\documentstyle[aps,preprint,prb]{revtex}
\begin{document}
\draft

\title{
Theory of phase-locking in generalized hybrid Josephson junction arrays}

\author{M. Basler\thanks{pmb@rz.uni-jena.de},
W. Krech\thanks{owk@rz.uni-jena.de} and
K. Yu. Platov\thanks{okp@rz.uni-jena.de}
}
\address{
\sl Friedrich-Schiller-Universit\"at Jena,\\
\sl Institut f\"ur Festk\"orperphysik,\\
\sl Max-Wien-Platz 1, D-07743 Jena, Germany}
\date{May 24, 1996}
%

\maketitle
\begin{abstract}
A recently proposed scheme for the analytical treatment of the dynamics of 
two-dimensional hybrid Josephson junction arrays is extended to a class of 
generalized hybrid arrays with ''horizontal'' shunts involving a 
capacitive as well as an inductive component. This class of 
arrays is of special interest, because the internal cell coupling 
has been shown numerically to favor in-phase synchronization for certain 
parameter values. As a result, we derive limits on the circuit 
design parameters for realizing this state. In addition, we obtain 
formulas for the flux-dependent frequency including flux-induced switching 
processes between the in-phase and anti-phase oscillation regime. The 
treatment covers unloaded arrays as well as arrays shunted via an 
external load.
\end{abstract}
\vskip2ex

\centerline{Submitted to {\it Phys. Rev. B}}
\vskip2ex


\section{Introduction}

Two-dimensional Josephson junction arrays are considered as strong 
candidates for tunable microwave oscillators. Since the pioneering 
works by Benz and Burroughs\cite{benz1,benz2} there were some attempts 
to fabricate arrays of this type\cite{booi1,sohn1,shea1} as well as to 
understand them theoretically 
\cite{wiesenfeld1,filatrella1,kautz1,darula1}. 
(For two recent reviews on 2D Josephson junction arrays, see Lachenmann 
\cite{lachenmann2} and Booi\cite{booi3}.) While radiation output of 
two-dimensional arrays should 
be much larger than that of one-dimensional arrays (for quadratic arrays 
in the matched case typically $\sim N^2$, with $N$=number of rows, 
compared to being $\sim N$, with $N$=number of junctions, for 
one-dimensional arrays\cite{wiesenfeld1,likharev1}) observations point 
more to the opposite direction. While in one-dimensional arrays there were 
observed up to 160$\mu$W,\cite{lukens2,booi3} the output power reported in 
two-dimensional 
arrays is several orders of magnitude smaller with a maximum of around 
400nW\cite{benz2,stern1,octavio1}.

Potentially, there can be several reasons responsible for this discrepancy. 
Besides low critical currents/normal resistances, mismatch to the external 
load, or parameter tolerances some more basic problems might 
be responsible for this. Indeed, some recent theoretical 
investigations show, that the radiating in-phase mode is neutrally stable 
in a unshunted array without external flux\cite{wiesenfeld1} and, even 
worse, that it 
is unstable even for a small flux entering the cell\cite{basler4}. As a 
result, the natural state of at least the simple model circuit studied 
in\cite{basler4} is a non-radiating one with both cells 
oscillating against each other. The situation can be improved by adding an 
appropriate external shunt synchronizing in-phase via its long-range 
interaction, but generally there remains a tendency that pairs of cells 
lock anti-phase and drop out of the radiating mode. 

A recently proposed layout\cite{krech14} removes this difficulty by 
introducing an additional capacitive shunt in the ''horizontal'' branches 
thus turning the internal coupling to favor the in-phase state. It is the 
aim of the present investigation, to give this idea an analytical 
foundation and derive some rigorous results, notably on the parameter 
boundaries separating in-phase from anti-phase oscillations. In addition, 
we study the interplay with an external load leading to a rather 
complex picture of possible stability regions as a result of the competition
of external and internal synchronization.

We start with an exposition of the problem including the basic equations 
in Sec. II. In Sec. III and IV these equations are solved by an analytical 
approximation, combining ideas of the strong coupling method appropriate 
for small-inductance Josephson junction cells\cite{basler3} with the 
standard 
weak-coupling procedure of slowly-varying phase\cite{jain1,krech6,krech7} 
for the treatment of inter-cell coupling. While Sec. III contains 
lowest-order results corresponding to vanishing cell inductance, Sec. IV 
includes the effects caused by a small, but non-vanishing inductance 
being essential for understanding the inter-cell coupling.
Sec. V contains several results including a comparison with numerical 
simulations. The interplay with an external load is treated in Sec. VI, and 
Sec. VII contains several more general conclusions relevant for the layout 
of two-dimensional Josephson junction 
arrays.

\section{The model and the basic equations}
For making the problem accessible to an analytical treatment, we have to 
make several propositions. Fig. 1 shows the circuit under 
consideration. To make the physical mechanisms more transparent,
the external shunt $Z_S$ is removed in the beginning, and 
will only be included in Sec. VI. Despite its simplicity this model has all 
the main features present in larger arrays, too: It is truly 
two-dimensional with a possible flux entering the cells. Notice, that 
unlike to conventional hybrid arrays\cite{shea1,kautz1} 
the horizontal branch contains a more general shunt consisting not only of 
the usual inductive connection, but of a parallel capacitance and 
resistance. Numerical results obtained before indicate that an intrinsic 
shunt of this type can favor in-phase locking even in an externally 
unshunted array. Here, we will confirm and extend this result by 
developing an analytical formalism which should be applicable to 
larger two-dimensional arrays with the same general structure as well.

Some more restrictions have to be put on the array: (i) Josephson junctions 
are described by the RSJ model\cite{likharev1}. (ii) All junctions are 
considered to be identical. (iii) Junctions are overdamped with a McCumber 
parameter set to zero. (iv) Self-inductance is taken into account while 
mutual inductance is neglected.  (v) The normalized ring inductance 
between the two loops
\begin{equation}\label{lsmall}
l=2\pi I_C L/\Phi_0
\end{equation}
is supposed to be small ($l\ll1$).
From the beginning, one has to understand that 
the inductance of the horizontal connection acts in a two-fold way. At 
first, it contributes to both ring inductances thus determining the SQUID 
coupling within each loop. At second, it is part of the shunt common to 
both loops and as such it influences the inter-cell coupling. With Eq. 
(\ref{lsmall}) we request the SQUID coupling to be strong, which is a 
necessary prerequisite for our approximation scheme to work. On the other 
hand, we will not fix the ratio between the inductive and the capacitive 
horizontal impedances from the beginning.

In the following, we will exploit some more normalized quantities as 
follows,
\begin{eqnarray}
s&=&\frac{2e}{\hbar}I_CR_Nt,\\
\varphi&=&2\pi\Phi/\Phi_0,\\
c&=&\frac{2e}{\hbar}I_CR_N^2C,\label{cnorm}\\
r&=&R/R_N,\label{rnorm}\\
i&=&I/I_C,
\end{eqnarray}
with $I_C$ the junction critical current, $R_N$ the normal resistance of 
one of the (identical) junctions, $\Phi$ the external flux per cell, and 
the last normalization being valid for all currents entering the 
calculation. Adopting these normalizations, the 
circuit can be described by the RSJ equations for the Josephson phases,
\begin{equation}
\dot{\phi}_{ij}+\sin\phi_{ij}=i_{ij},\quad (\{i,j\}=\{1,2\})
\end{equation}
in conjunction with the two flux quantization conditions
\begin{equation}\label{fluxquan}
\phi_{i2}-\phi_{i1}-\varphi \mp li_l=0
\end{equation}
(minus sign refers to $i=1$) and Kirchhoff's current laws
\begin{eqnarray}
i_0&=&\frac{1}{2}(i_{11}+i_{12}),\\
\overline{i}&=&i_{11}-i_{21}=i_{22}-i_{12},\\
i_l&=&\overline{i}+i_{rc}.
\end{eqnarray}
These have to be supplemented by Kirchhoff's voltage law
\begin{equation}\label{kirchhoff2}
\ddot{i}_l+\frac{r}{l}\dot{i}_{rc}+\frac{1}{lc}i_{rc}=0.
\end{equation}
We would like to point out that while the inductive branch carrying 
current $i_l$ is part of both superconducting loops thus contributing to 
the flux quantization conditions (\ref{fluxquan}), the branch $i_{rc}$ 
enters only via the ordinary Kirchhoff's law (\ref{kirchhoff2}). As a 
result, it is impossible to 
simply substitute the three elements $l,c$ and $r$ by a single impedance 
{Z} from the beginning.

Before, it has proven useful in the treatment of strongly coupled SQUID 
cells\cite{basler3} to combine the Josephson phases within each cell via
\begin{eqnarray}\label{sigmadelta}
\Sigma_{k}&=&\frac{1}{2}\left(\phi_{k2}+\phi_{k1}\right),\\
\Delta_{k}&=&\frac{1}{2}\left(\phi_{k2}-\phi_{k1}\right).
\end{eqnarray}
In addition, we introduce the circular currents
\begin{equation}\label{circular}
i_k^{\circ}=(i_{k2}-i_{k1})/2.
\end{equation}
With the help of Eqs. (\ref{sigmadelta}) -- (\ref{circular}) we finally 
obtain the system
\begin{mathletters}
\label{basic}
\begin{eqnarray}
\dot{\Sigma}_k+\sin\Sigma_k\cos\Delta_k&=&i_0,\label{basic1}\\
\dot{\Delta}_k+\sin\Delta_k\cos\Sigma_k&=&i_k^{\circ},\label{basic2}\\
\Delta_1+\Delta_2-\varphi&=&0,\label{basic3}\\
\Delta_1-\Delta_2=li_l&=&l(i_2^{\circ}-i_1^{\circ}+i_{rc}),\label{basic4}\\
\ddot{i}_{rc}+\frac{r}{l}\dot{i}_{rc}+\frac{1}{lc}i_{rc}&=&
(\ddot{i_1}^{\circ}-\ddot{i_2}^{\circ})\label{basic5}
\end{eqnarray}
which our analytical approximation scheme is based on. As there are seven 
equations for the seven variables $\Sigma_k,\Delta_k,i_k^{\circ},i_{rc},$
this is a well-posed problem.

\end{mathletters}

\section{Analytical approximation scheme and lowest order results}
Our strategy for solving system (\ref{basic}) will be based on a 
perturbative treatment valid for small $l$ (for the basic idea compare our 
earlier paper\cite{basler3}). Thus, we start solving Eqs. (\ref{basic}) 
for $l=0$, and only later include corrections $\sim l$ exploiting the 
lowest order results obtained before. This procedure is favored by the 
fact that $l$ enters equation (\ref{basic4}) only. We start by 
evaluating (\ref{basic4}) in conjunction with (\ref{basic3}). The 
solutions for 
$\Delta_k$ can be used to evaluate $\Sigma_k$ from (\ref{basic1}). Next, 
we find the $i_k^{\circ}$ (not the $\Delta_k$, which are already known in 
this order!) from (\ref{basic2}). Finally, with the ring currents 
$i_k^{\circ}$ on the right hand side of (\ref{basic5}) known we can 
evaluate the current $i_{rc}$ by 
solving the corresponding differential equation. All 
other quantities, like $i_l$ or $\overline{i}$, are secondary and can be 
derived 
from the seven variables mentioned so far. Afterwards, we insert the 
lowest order 
result on the right hand side of Eq. (\ref{basic4}) and start a second 
cycle in the 
same sequence. 

The procedure described above gives the following lowest-order results. 
First, the Josephson phase differences in both loops are found to be 
identical,
\begin{equation}\label{delta0}
\Delta_{k,0}=\varphi/2.
\end{equation}
In the following, comma-delimited indices refer to the order of 
approximation. From (\ref{basic1}), the Josephson phases are found to 
coincide with the corresponding solutions for an autonomous 
junction,\cite{likharev1}
\begin{equation}\label{sigma0}
\Sigma_{k,0}=\frac{\pi}{2}+2\arctan\left(\frac{\zeta_0}{i_0+\cos(\varphi/2)}
\tan\frac{\zeta_0 s-\delta_k}{2}\right),
\end{equation}
with the important modification that the frequency $\zeta_0$ becomes 
flux-dependent according to
\begin{equation}
\zeta_0=\sqrt{i_0^2-\cos^2(\varphi/2)}.
\end{equation}

Next, the circular currents can be evaluated from (\ref{basic2}). Note, 
that this equation originating from the original Josephson equations does 
not lead to a differential equation, because the constant Josephson phase 
differences $\Delta_k$ are already known. The result is
\begin{equation}\label{circ0}
i_{k}^{\circ}=\sin(\varphi/2)\cos\Sigma_{k,0}.
\end{equation}
It is a trivial task to evaluate Eq. (\ref{circ0}) using (\ref{sigma0}); in 
the further calculation we will only need the lowest harmonics of the 
circular currents,
\begin{equation}
i_{k,0}^{\circ}=-2\frac{\zeta_0}{i_0+\zeta_0}\sin(\varphi/2)
\sin(\zeta_0 s-\delta_k).
\end{equation}
The corresponding difference of the ring currents,
\begin{equation}\label{ibar0}
\overline{i}_{,0}=i_{2,0}^{\circ}-i_{1,0}^{\circ}=
4\frac{\zeta_0}{i_0+\zeta_0}\sin(\varphi/2)\sin\left(\frac{\delta_1-
\delta_2}{2}\right)
\cos\left(\zeta_0 s-\frac{\delta_1+\delta_2}{2}\right),\label{ibar}
\end{equation}
enters the horizontal connection thus acting as a driving force for 
the oscillatory circuit according to Eq. (\ref{basic5}). This equation can 
be solved with standard methods. The stationary oscillating solution reads
\begin{equation}
i_{rc,0}=-\frac{4l\zeta_0^2}{|Z(\zeta_0)|(i_0+\zeta_0)}\sin(\varphi/2)
\sin\left(\frac{
\delta_1-\delta_2}{2}\right)\sin\left(\zeta_0 
s-\frac{\delta_1+\delta_2}{2}-\psi(\zeta_0)\right).\label{irc}
\end{equation}
Here, we introduced the series circuit impedance $Z$ with
\begin{equation}
|Z(\zeta_0)|=\sqrt{r^2+\left(\frac{1}{c\zeta_0}-l\zeta_0\right)^2}
\end{equation}
and the phase angle $\psi$,
\begin{equation}
\cos\psi(\zeta_0)=\frac{r}{|Z(\zeta_0)|},\qquad
\sin\psi(\zeta_0)=\frac{l\zeta_0-\displaystyle\frac{1}{c\zeta_0}}{
|Z(\zeta_0)|}.
\end{equation}

For later purposes we need $i_{l}=\overline{i}+i_{rc}$
rather than $i_{rc}$, because it is just $i_l$ which 
potentially may split the oscillation phases between cell 1 and 
cell 2 via Eq. (\ref{basic4}). Combining (\ref{irc}) with (\ref{ibar}) 
after some algebra we obtain
\begin{equation}\label{il0}
i_{l,0}=-\frac{4\zeta_0}{i_0+\zeta_0}\frac{|z(\zeta_0)|}{
|Z(\zeta_0)|}\sin(\varphi/2)\sin
\frac{\delta_1-\delta_2}{2}\cos\left(\zeta_0 s-\frac{\delta_1+\delta_2}{2}-
\chi(\zeta_0)\right),
\end{equation}
where we introduced the $rc$ impedance $|z|$ with
\begin{equation}
|z(\zeta_0)|=\sqrt{r^2+\frac{1}{(c \zeta_0)^2}}
\end{equation}
and 
\begin{equation}
\sin\chi(\zeta_0)=\frac{rlc\zeta_0^2}{|Z(\zeta_0)
|\sqrt{1+(rc\zeta_0)^2}},\qquad
\cos\chi(\zeta_0)=\frac{r^2c\zeta_0+\left(\displaystyle
\frac{1}{c\zeta_0}-l\zeta_0\right)    
}{|Z(\zeta_0)|\sqrt{1+(rc\zeta_0)^2}}.
\end{equation}
(In principle, one could evaluate $i_l$ directly from an equation similar 
to (\ref{basic5}) of course, and we checked that the result is the same. 
The procedure described here has the advantage of additionally providing 
an expression for the current flowing through the capacitive line.)

To summarize, we observe the following lowest-order results: All four 
junctions oscillate with the same flux-dependent frequency 
$\zeta_0=\sqrt{i_0^2-\cos^2(\varphi/2)}$. Because of 
(\ref{delta0}), the junctions within each cell are exactly in phase, while 
the relative phase between cell 1 and cell 2 (given by $\delta_1$ and 
$\delta_2$, respectively) is undetermined, up to now. If both cells are in 
phase, 
there is no current through the horizontal line, because of the 
$\sin[(\delta_1-\delta_2)/2]$ present in (\ref{ibar0}). On the other hand, 
the horizontal current reaches its maximum if both cells oscillate 
anti-phase with 
$\delta_1-\delta_2=\pi$.

\section{Inductance effects}
Now we are ready to include inductance effects. Again starting with 
(\ref{basic3}) and (\ref{basic4}), we insert the lowest-order result 
(\ref{il0}) on the right hand side of (\ref{basic4}). This leads to 
\begin{eqnarray}\label{delta12}
\Delta_{1/2}&=&\Delta_{1/2,0}+l\Delta_{1/2,1}\\
&=&\frac{\varphi}{2}\pm2l
\frac{\zeta_0}{i_0+\zeta_0}\frac{|z|}{|Z|}\sin(\varphi/2)\sin
\frac{\delta_2-\delta_1}{2}
\cos\left(
\zeta_0 s-\frac{\delta_1+\delta_2}{2}-\chi\right).\nonumber
\end{eqnarray}
Note, that the first index in (\ref{delta12}) refers to cell 1 and cell 2, 
respectively, while the second one indicates the order of evaluation; the 
$+$ sign refers to $\Delta_1$. This has to be inserted into (\ref{basic1}),
\begin{equation}\label{sigmak1}
\dot{\Sigma}_{k}+\cos(\Delta_{k,0}+l\Delta_{k,1})\sin\Sigma_{k}=i_0.
\end{equation}
For evaluating these equations the cosine on the left hand side is expanded 
according to
\begin{equation}\label{deltadot}
\cos(\Delta_{k,0}+l\Delta_{k,1})\approx 
\cos\Delta_{k,0}-l\Delta_{k,1}\sin\Delta_{k,0}.
\end{equation}
After transferring the correction term $\sim l$ to the right hand side of
(\ref{sigmak1}) one makes the crucial  observation, that it acts in a 
similar way as,for example, an external shunt synchronizing the 
cells\cite{jain1}. 

The resulting equations are evaluated with the conventional phase-slip 
method (see, for instance\cite{jain1,krech6,krech7}). According to this 
procedure which has proven useful in the treatment of linear arrays 
before, the up to now constant phases $\delta_1$ and $\delta_2$ are 
considered as time dependent,
\begin{equation}
\delta_{k}=\delta_{k}(s),
\end{equation}
with the subsidiary condition that this time-dependence is only an 
adiabatic one,
\begin{equation}
\dot{\delta}\ll\zeta_0.
\end{equation}
Physically, this means that the phases are required to be nearly constant 
during one Josephson oscillation.

With these assumptions, the same ansatz (\ref{sigma0}) with $\delta_k(s)$ 
and $\zeta$ instead of $\zeta_0$ leads to the sum voltages $\dot{\Sigma}_k$,
\begin{equation}\label{sigma1}
\dot{\Sigma}_k=\frac{\zeta_0(\zeta-\dot{\delta}_k)}{i_0+
\cos(\varphi/2)\cos(\zeta s-\delta_k)}.
\end{equation}
Writing $\zeta$ instead of $\zeta_0$ we have allowed for a possible 
(small) deviation of the actual oscillation frequency from $\zeta_0$.
Inserting (\ref{sigma1}) into (\ref{sigmak1}) leads to the reduced equations
\begin{equation}\label{phaseslip}
\zeta_0(\zeta-\zeta_0-\dot{\delta}_k)=l\sin(\varphi/2)\Delta_{k,1}
\left(\cos(\varphi/2)+i_0\cos (\zeta s-\delta_k)\right).
\end{equation}

After averaging over one time period and applying some algebra we arrive 
at the following system of equations (for details see, for 
instance\cite{basler4,jain1,krech6,krech7})
\begin{mathletters}
\label{delta1-2}
\begin{eqnarray}
\zeta_0(\zeta-\zeta_0-<\dot{\delta}_1>)&=&
li_0
\frac{\zeta_0}{i_0+\zeta_0}\frac{|z|}{|Z|}\sin(\varphi/2)\sin
\frac{<\delta_2>-<\delta_1>}{2}
\sin(\varphi/2)\nonumber\\
&&\times\cos\left(
\frac{<\delta_2>-<\delta_1>}{2}+\chi
\right),\label{delta1}\\
\zeta_0(\zeta-\zeta_0-<\dot{\delta}_2>)&=&
-li_0
\frac{\zeta_0}{i_0+\zeta_0}\frac{|z|}{|Z|}\sin(\varphi/2)\sin
\frac{<\delta_2>-<\delta_1>}{2}
\sin(\varphi/2)\nonumber\\
&&\times\cos\left(
\frac{<\delta_2>-<\delta_1>}{2}+\chi
\right),\label{delta2}
\end{eqnarray}
\end{mathletters}

\noindent
where $<>$ denotes the time average over one Josephson oscillation.
The difference of (\ref{delta1}) and (\ref{delta2}) gives an evolution 
equation for the phase difference $<\delta>$,
\begin{equation}\label{deltasol}
<\dot{\delta}>=\frac{i_0l}{i_0+\zeta_0}\frac{|z|}{|Z|}\sin^2(\varphi/2)
\cos\chi\sin<\delta>.
\end{equation}
Eq. (\ref{deltasol}) is the basic equation determining the possible phase 
differences between the oscillations of both cells as well as the 
corresponding 
regions of stability.

\section{Phase locking, stability and oscillation frequency}
We will not go into the question of general solutions of (\ref{deltasol}) 
but concentrate on phase-locking, being characterized by a 
time-independent phase-shift between cell 1 and cell 2,
\begin{equation}\label{deltadot1}
<\dot{\delta}^{\mathrm{lock}}>=0.
\end{equation}
Within the range $0\le\delta<2\pi$ there are obviously only two 
possibilities for Eq. (\ref{deltadot1}) to be valid,
\begin{equation}
<\delta^{\mathrm lock}>=0\quad\mbox{and}\quad<\delta^{\mathrm lock}>=\pi,
\end{equation}
the first one describing in-phase oscillations and the second one 
anti-phase oscillations of the cells.

The crucial question of the range of stability of these two solutions can 
be answered on the basis of the evolution equation (\ref{deltasol}), too. 
The ansatz
\begin{equation}
<\delta>=<\delta^{\mathrm lock}>+a\mbox{e}^{\lambda t}
\end{equation}
($|a|\ll|1|$) leads to the Lyapunov coefficient
\begin{equation}
\lambda=\frac{i_0l}{i_0+\zeta_0}\frac{|z|}{|Z|}\sin^2(\varphi/2)\cos\chi
\cos<\delta^{\mathrm lock}>.
\end{equation}
One recovers, that the stability is solely determined by the $\cos\chi$;
all the remaining factors, except $\delta^{\mathrm lock}$, are positive 
definite. In detail, the
\begin{equation}\label{inanticos1}
\mbox{in-phase solution $<\delta^{\mathrm lock}>=0$ is stable for 
$\cos\chi<0$,}
\end{equation}
while the
\begin{equation}\label{inanticos2}
\mbox{anti-phase solution $<\delta^{\mathrm lock}>=\pi$ is stable for 
$\cos\chi>0$.}
\end{equation}

Before further evaluating this condition we will consider the 
oscillation frequency which can be derived from (\ref{delta1}) (or 
(\ref{delta2})). With
\begin{equation}
<\delta_1>=<\delta_2>=\mbox{const}=0
\end{equation}
one easily recovers
\begin{equation}\label{zetain}
\zeta^{\mathrm{in}}=\zeta_0=\sqrt{i_0^2-\cos^2(\varphi/2)}.
\end{equation}
Evaluating the anti-phase frequency with
\begin{equation}
<\delta_1>-<\delta_2>=\pi
\end{equation}
needs a bit more algebra. The result is
\begin{equation}\label{zetaanti}
\zeta^{\mathrm{anti}}=\zeta_0\left(
1-\frac{i_0l^2r\sin^2(\varphi/2)}{|Z|^2(i_0+\zeta_0)}\right).
\end{equation}
Thus, if both cells oscillate in-phase their frequency is identical to the 
autonomous oscillation frequency. On the other hand, if the cells 
oscillate anti-phase the frequency will be lower than $\zeta_0$.
The physical reason for this behavior can be understood by comparing with 
other (even linearly) oscillating systems: If the bindings (in our case 
realized by the horizontal impedance) are not loaded, the oscillation 
frequency remains the same as for uncoupled oscillators; if the 
bindings are loaded (i.e. in case of a ac current flowing through the 
horizontal line) the system oscillates with a different frequency.

Unfortunately, one has to respect a certain limit of validity of Eq. 
(\ref{zetaanti}). Using the method of slowly varying phase we have adopted 
the supposition mentioned before that the frequency must not deviate too 
much from $\zeta_0$,
\begin{equation}
\zeta\approx\zeta_0.
\end{equation}
Thus, the correction in Eq. (\ref{zetaanti}) is required to be small 
compared to the frequency itself. A rough estimate valid for $i_0>1.15$ 
leads to the 
condition
\begin{equation}
l^2\ll r.
\end{equation}
Our experience shows that usually a factor of $2\ldots3$ is sufficient for 
this condition to be fulfilled.

Now we return to the question of anti-phase $\leftrightarrow$ in-phase 
transitions described by (\ref{inanticos1}) and (\ref{inanticos2}), resp.
Considering the numerator of $\cos\chi$ one observes that the boundary 
separating in-phase and anti-phase oscillations of the cells is given by
\begin{eqnarray}\label{transgen}
\left(\frac{1}{c\zeta}-l\zeta\right)+r^2c\zeta=0
\end{eqnarray}
with the cells oscillating anti-phase if the left hand side is positive and 
in-phase if it is negative. In other words, the transition between both 
regimes lies in the vicinity of the resonance curve of the $l$-$c$-$r$ 
connection with deviations becoming important for small $l$. Fig. 2 shows 
the boundary between the two regimes for a frequency $\zeta=1.11$ in 
comparison to numerical results.

To summarize, the in-phase regime is favored for not too large $r$ as long 
as the inductive impedance dominates over the capacitive one, while for 
the capacitive impedance dominating the cells oscillate anti-phase. There 
is a simple physical explanation for this: Anti-phase oscillations are 
caused by the flux coupling via the joint inductive line carrying current 
$i_l$. For a sufficiently large capacitive shunt, the current prefers the 
capacitive way
which does not produce any such flux.

In conventional hybrid arrays\cite{kautz1} horizontal lines are purely 
inductive. Formally, this limit can be observed letting $c\to0$. In this 
case the capacitive impedance goes to infinity while the correction $\sim 
r^2c$ tends to zero. Then, there is no possibility for the current 
to be shunted, and the cells remain in the anti-phase regime\cite{basler4}.
The more general question, for which parameter values $l,c$, and $r$ there 
are no transitions can be answered on the basis of Eq. (\ref{transgen}). 
This equation does only have real solutions for $\zeta$ if
\begin{equation}
l>r^2c.
\end{equation}
For all smaller $l$, the current in the inductive line is strong enough to 
keep the cells oscillating anti-phase.

Considering the circuit parameters $i_0, l$ etc. as constant and leaving 
the external flux $\varphi$ as the only free parameter one can observe 
flux-induced transitions between both regimes. The difference between the 
frequencies $\zeta^{\mathrm in}$ and $\zeta^{\mathrm anti}$ leads to a 
hysteresis, which has been observed in numerical simulations 
before\cite{krech14}. In more detail, in-phase $\to$ anti-phase 
transitions are observed at
\begin{equation}\label{phiinanti}
\varphi^{\mathrm{ia}}=2\arccos\left[\pm\sqrt{i_0^2-\zeta^{{\mathrm tr}2}}
\right],
\end{equation}
where we introduced the transition frequency
\begin{equation}\label{transfreq}
\zeta^{\mathrm tr}=\frac{1}{\sqrt{lc-r^2c^2}}
\end{equation}
as can be easily deduced from (\ref{transgen}). The transition from the 
anti-phase to the in-phase regime needs a bit more algebra. It can be 
determined from the requirement, that the anti-phase frequency 
(\ref{zetaanti}) be equal to the transition frequency 
(\ref{transfreq}). Unfortunately, the resulting equation can not be solved 
in closed form. However, as a first approximation, one can equate the 
in$\to$anti transition frequency (\ref{transfreq}) with (\ref{zetaanti})
and evaluate for $\varphi$, substituting $\zeta\to\zeta^{\mathrm tr}$ on 
the right hand side of (\ref{zetaanti}),
\begin{equation}
\varphi^{\mathrm{ai}}=2\arccos\left(
\pm\sqrt{\frac{i_0^2-(\zeta^{\mathrm tr})^2-\frac{\displaystyle 
2i_0l}{\displaystyle cr\left(
i_0+\zeta^{\mathrm tr}\right)}}{1-\frac{\displaystyle 2i_0l}{\displaystyle 
cr\left(i_0+\zeta^{\mathrm tr}\right)}}
}
\right).
\end{equation}
It can be deduced, that $\varphi^{\mathrm{ai}}$ is always larger than 
$\varphi^{\mathrm{ia}}$. A better result for $\varphi^{\mathrm{ai}}$ is 
obtained by graphically finding the transition frequency on the curve at 
$\zeta=\zeta^{\mathrm tr}$.

Thus, if there are any transitions between both regimes at all, for small 
values of the external flux the cells oscillate with the lower anti-phase 
frequency switching to in-phase oscillations at $\varphi^{\mathrm ai}$. 
Because of $\varphi^{\mathrm ai}>\varphi^{\mathrm ia}$ (for 
$0<\varphi\le\pi$) switching back to the anti-phase state occurs at 
a lower flux, leading to the hysteresis mentioned above. Fig. 3 shows a 
plot of frequency against flux in comparison with the outcome of a 
numerical simulation. The frequencies are in excellent agreement, and even 
the transition points, which depend rather sensibly on the parameters, are 
located within the same region.

This last result concerning hysteresis has to be taken with some care. It 
was obtained by combining the anti-phase frequency formula 
(\ref{zetaanti}) with Eq. 
(\ref{transgen}) and evaluating for $\varphi$. However, (\ref{transgen}) 
as originating from (\ref{deltasol}) is already a first order result, thus 
inserting (\ref{zetaanti}) might not be fully justified while second order 
terms in (\ref{deltasol}) are neglected. Nonetheless, it gives a plausible
explanation for the mechanism causing the hysteresis observed in 
numerical simulations.

\section{Long-range synchronization via an external load}
It has been well-known for a long time that synchronization in a 
one-dimensional array can be achieved and controlled by shunting the array 
via an external load\cite{jain1,krech5}. In a similar manner one may hope 
to be able to control row locking in two-dimensional arrays, too. For 
studying this mechanism within our model we now add the external load 
already indicated in Fig. 1. As a result, we have to supplement the basic 
equations (\ref{basic}). At first, we add the mesh rule for the load 
current $i_L$,
\begin{equation}\label{loadcurr}
\sum_{k=1,2}\ddot{\Sigma}_k-l_L\ddot{i}_L-r_L\dot{i}_l-\frac{1}{c_L}i_L=0.
\end{equation}
Here, $r_L,l_L,$ and $c_L$ are the load impedances normalized in the same 
manner as (\ref{lsmall}), (\ref{cnorm}), and (\ref{rnorm}).
In addition, the load current couples back to the junctions, thus 
supplementing Eq. (\ref{basic1}),
\begin{equation}\label{sumnew}
\dot{\Sigma}_k+\sin\Sigma_k\cos\Delta_k=i_0-\frac{1}{2}i_L.
\end{equation}

As has been observed in the study of similar one-dimensional 
synchronization problems before, the reciprocal impedance $1/|Z_L|\ll1$ 
provides another perturbation 
parameter for a sufficiently large load; thus we evaluate the system 
perturbatively, neglecting terms $\sim l/|Z_L|$. To lowest order with 
respect to 
$|Z_L|$ the load current vanishes, and we end up with the results 
described in 
Sec. III. Based on the lowest order 
Josephson oscillations (\ref{sigma0}) and the corresponding voltages 
$\dot{\Sigma}_{k,0}$ we obtain the first order (with respect to $1/|Z_L|$) 
load current $i_{L,0}$,
\begin{equation}\label{loadcurrsol}
i_{L,0}=\frac{4\cos(\varphi/2)}{|Z_L|}\frac{\zeta_0}{i_0+\zeta_0}\cos\left(
\frac{\delta_1-\delta_2}{2}\right)\sin\left(\zeta_0 
s-\frac{\delta_1+\delta_2}{2}-\psi_L\right)
\end{equation}
with
\begin{mathletters}
\begin{eqnarray}
|Z_L(\zeta_0)|&=&\sqrt{(r_L+1)^2+\left(\frac{1}{c_L\zeta_0}-l_L\zeta_0
\right)^2},\\
\sin\psi_L(\zeta_0)&=&\frac{r_L+1}{|Z_L(\zeta_0)|},\\
\cos\psi_L(\zeta_0)&=&\frac{\displaystyle\frac{1}{c_L\zeta_0}-l_L\zeta_0}{
|Z_L(\zeta_0)|}.
\end{eqnarray}
\end{mathletters}
Its structure is obviously quite similar to that of the horizontal 
current (\ref{ibar0}). However, one should note two differences: (i) While 
the load current is maximal for $\varphi=0$, the horizontal current 
reaches its maximum for $\varphi=\pi/2$. (ii) The horizontal current 
vanishes if both cells oscillate in-phase, while the load current vanishes 
for both cells oscillating anti-phase. 

The load current (\ref{loadcurrsol}) provides the additional 
contribution to (\ref{sumnew}) and, as a result, the phase slip equations
(\ref{delta1-2}) get an additional term, too. After performing the 
time-averages we get
\begin{mathletters}
\label{newdeltas}
\begin{eqnarray}
&&\begin{array}{rl}
\zeta_0(\zeta-\zeta_0-<\dot{\delta}_1>)&=-\displaystyle\frac{i_0l}{2}
\displaystyle\frac{
\zeta_0}{i_0+\zeta_0}\displaystyle\frac{|z|}{|Z|}\sin^2(\varphi/2)\left(
\sin\psi+\sin(<\delta_1>-<\delta_2>-\chi)\right)\\
&+\displaystyle\frac{1}{2}\displaystyle\frac{\zeta_0}{i_0+\zeta_0}
\frac{1}{|Z_L|}
\cos^2(\varphi/2)
\left(\sin\psi_L-\sin(<\delta_1>-<\delta_2>-\psi_L)\right),
\end{array}\label{newdeltas1}
\\
&&\begin{array}{rl}
\zeta_0(\zeta-\zeta_0-<\dot{\delta}_2>)&=\displaystyle\frac{i_0l}{2}\frac{
\zeta_0}{i_0+\zeta_0}\displaystyle\frac{|z|}{|Z|}\sin^2(\varphi/2)\left(
\sin\psi+\sin(<\delta_1>-<\delta_2>-\chi)\right)\\
&+\displaystyle\frac{1}{2}\displaystyle\frac{\zeta_0}{i_0+\zeta_0}
\frac{1}{|Z_L|}
\cos^2(\varphi/2)
\left(\sin\psi_L+\sin(<\delta_1>-<\delta_2>-\psi_L)\right).
\end{array}\label{newdeltas2}
\end{eqnarray}
\end{mathletters}

\noindent
By subtracting (\ref{newdeltas1}) and (\ref{newdeltas2}), we finally get 
the evolution equation for the averaged oscillation phase difference,
\begin{equation}
<\dot{\delta}>=\frac{1}{i_0+\zeta_0}\left(
\frac{1}{|Z_L|}\cos^2(\varphi/2)\cos\psi_L+
i_0l\frac{|z|}{|Z|}\sin^2(\varphi/2)\cos\chi
\right)\sin<\delta>.
\end{equation}
Despite the relatively complicated interplay between cell interaction via 
the horizontal line and long range coupling via the external load there 
remain only the same two phase locking solutions as before,
\begin{equation}
<\delta^{\mathrm lock}>=0\quad\mbox{and}\quad<\delta^{\mathrm lock}>=\pi,
\end{equation}
the stability of which is determined by the Lyapunov coefficient
\begin{equation}
\lambda=\frac{1}{i_0+\zeta_0}\left(
\frac{1}{|Z_L|}\cos^2(\varphi/2)\cos\psi_L+
i_0l\frac{|z|}{|Z|}\sin^2(\varphi/2)\cos\chi
\right)\cos<\delta^{\mathrm lock}>.
\end{equation}

In-phase oscillations of the cells are stable if the term in parenthesis is 
lower than zero while anti-phase oscillation are stable if it is greater 
than zero. Thus, the desired stability for the in-phase mode is 
reached for
\begin{equation}\label{stabload}
\frac{1}{|Z_L|}\cos^2(\varphi/2)\cos\psi_L+
i_0l\frac{|z|}{|Z|}\sin^2(\varphi/2)\cos\chi<0.
\end{equation}
Eq. (\ref{stabload}) shows a rather complex parameter dependence, relating 
the seven parameters $r, l, c, r_L, l_L, c_L,$ and $\varphi$.
Its physical meaning is best discovered considering several limiting cases.

(i) For a sufficiently large external load,
\begin{equation}
\frac{1}{|Z_L|}\ll i_0l\frac{|z|}{|Z|},
\end{equation}
the relative phase of the cells is determined by the internal coupling 
alone. This has to be compared to the case of two externally loaded 
separate cells\cite{basler5}. In this case -- as for linear arrays -- the 
relative phase depends on the character of the external load only: While 
for inductively dominated loads the cells are locked in-phase, they are 
locked anti-phase for capacitively dominated loads, independently of the 
magnitude of the external load.

(ii) The contributions from the external load and from the internal shunt 
show a different flux dependence. For sufficiently small values of 
external flux the last term can be neglected, and the locking regime is 
controlled by the load only. On the other hand, for flux values of around 
half a flux quantum the first term becomes negligible, and the internal 
horizontal line determines the phase difference of the cells.

(iii) For $l\to0$, the second term can be neglected, and the result agrees 
with that obtained for two separate cells before\cite{basler5}, as it 
should be. In this limit the cells internally decouple, while the external 
coupling remains in force.

(iv) The usual hybrid arrays without the internal $R$-$C$-line are 
contained as a limiting case. For $r\to\infty$, the in-phase-condition, 
Eq. (\ref{stabload}), reduces to
\begin{equation}\label{hybridinanti}
\frac{1}{|Z_L|}\cos\psi_L\cos(\varphi^2/2)+i_0l\sin^2(\varphi/2)<0.
\end{equation}
It states, that for sufficiently large inductances,
\begin{equation}
l>l^{\mathrm cr}=-\frac{\cos\psi_L}{i_0\tan^2(\varphi/2)|Z_L|},
\end{equation}
ordinary pure inductive hybrid arrays may switch to the anti-phase state 
even for inductive external loads.

The indicated transition was indeed observed in a numerical simulation 
(boxes in Fig. 4). Having in mind that Eq. (\ref{hybridinanti}) is the 
result of several approximations, concerning the external shunt as well as 
the internal inductive coupling, the agreement is remarkably good.

The influence of changing parameters can be nicely illustrated
by performing a second simulation with exactly the same parameter set, but 
distributing ring inductance $l$ regularly around the loops. 
The result denoted by the dots in Fig. 4 clearly deviates from that 
obtained for inductance concentrated on the horizontal line considered 
before. This can be taken as a strong indication that the coupling is not 
provided by the loop inductances but by the inductance on the line 
common to both cells.

\section{Conclusions}
Although our work is devoted to the study of a simple model circuit
several results are expected to be valid for larger arrays, too. At first, 
the short range coupling between neighboring cells leads to an anti-phase 
synchronization in conventional Josephson junction hybrid arrays. This may 
be one reason for explaining the very small radiation output in 2D 
Josephson junction arrays obtained so far. At second, we show a way to 
improve the situation by adding a capacitive shunt parallel to the 
horizontal lines. In this way, the flux-generating current potentially 
responsible for the anti-phase coupling is redirected through the 
capacitive line which is not part of a flux quantization condition.

Combining Fig. 2 with some already known facts on synchronization in 
strongly coupled SQUID cells\cite{basler3} the following design criteria 
for generalized hybrid 2D Josephson junction arrays can be derived. (i) 
For synchronizing horizontal 
lines in-phase the ring inductances have to be kept small ($l\ll1$). (ii) 
In-phase synchronization between neighboring cells in vertical direction 
is observed for $l>1/c\zeta^2+r^2c$. Based on this, we will derive 
some estimates for reasonable $c$ and $l$. For a given $l$, the boundary 
between in- and anti-phase oscillations is given by Eq. (\ref{transgen}). 
Fig. 2 shows already that the additional term $\sim r$ restricts the 
possible $l$ by setting a lower bound. This bound is obtained from
\begin{equation}
\frac{\mbox{ d}l}{\mbox{ d}c}=0
\end{equation}
as
\begin{equation}\label{cr}
r=\frac{1}{c\zeta}\quad\mbox{resp.}\quad l=2r/\zeta.
\end{equation}
Thus, for obtaining in-phase oscillations the condition
\begin{equation}
r<\frac{\zeta l}{2}
\end{equation}
has to be respected. Because of (\ref{cr}) this means
\begin{equation}
c>\frac{2}{l\zeta}.
\end{equation}
Obviously, the requirement to have a small $l$ for horizontal in-phase 
synchronization leads to the demand to have a sufficiently high 
capacitance $c>2/l\zeta$ as well as a small resistance 
$r<l\zeta/2$.
A reasonable compromise might for instance be
\begin{equation}
l\approx0.8\qquad c\approx 3.0\qquad r\approx 0.2.
\end{equation}

Of cause, all these estimates should be considered as very rough, and on 
the other hand, one has to check carefully how large these quantities 
on chip actually are.

On the other hand, we would like to point out that these suggestions are 
based on an analytical approximation scheme and are founded on solid 
formulae. Of cause, it still has to be shown rigorously that they can be 
transferred to larger arrays as well. Some preliminary results from 
numerical simulations indeed indicate this. We hope, that the general 
procedure described here can be transferred to larger arrays of the type 
considered here as well, and some work is on the way to actually extend it 
to a ladder configuration.

If the arrays are externally loaded, which is usually done via an 
inductive load, the parameters have to be chosen in such a manner to 
respect Eq. (\ref{stabload}). The best way for obtaining in-phase 
synchronization is to make both contributions to the Lyapunov-coefficient 
lower than zero separately, which is possible because the parameters of 
the external load can be chosen independently of those from the internal 
shunt. In general, one should select values such, that (i) the external 
load is dominated by its inductive contribution, (ii) the internal 
horizontal shunts are dominated by the inductive impedance, too. Because 
of the frequency-dependence of the characters of the shunts, one has to 
make sure, that these conditions are met for all values of external flux.

Of cause, the circuit studied here has several features requiring a more 
detailed investigation, either analytically or numerically. Usually one 
exploits shunted tunnel junctions for building arrays, thus one may ask 
for the influence of non-vanishing McCumber parameters. On the other 
hand, the influence of parameter splitting needs to be investigated, and 
in addition, in real arrays, noise comes into play. While this last aspect 
is to be expected to play only a minor role within the small inductance 
loops, it will be sure have some influence on the coupling between the 
cells.

\section*{Acknowledgments}
This work was supported by a project of the Deutsche 
Forschungsgemeinschaft DFG under contract \# Kr1172/4-1. The authors would 
like to express their thanks to the DFG for financial support.
\eject

\begin{figure}
\caption{The generalized Josephson junction hybrid array model circuit 
under investigation.~~~~~~~~~~~~~~~~~~}
\label{fig1}
\end{figure}

\begin{figure}
\caption{The boundary between in-phase and anti-phase oscillations. Solid 
line: analytical approximation. Crosses: numerical simulation. 
Parameters: $i_0=1.5,r=0.1,\varphi=1.0$.}
\label{fig2}
\end{figure}

\begin{figure}
\caption{Frequency against flux with a transition from anti-phase to 
in-phase oscillations. Parameters: $i_0=1.5, r=0.1, l=0.2, c=4.0$. (a) 
Analytical approximation. (b) Numerical simulation.}
\label{fig3}
\end{figure}

\begin{figure}
\caption{Transition between in-phase and anti-phase state caused by the 
internal inductive coupling present in a hybrid array. 
Parameters: $i_0=1.5, r_L=1.0, l_L=1.0, c_L=2.0$. Solid line: analytical 
approximation, boxes: numerical simulation, open dots: numerical 
simulation with inductance regularly distributed around loops.}
\label{fig4}
\end{figure}


\bibliographystyle{prsty}

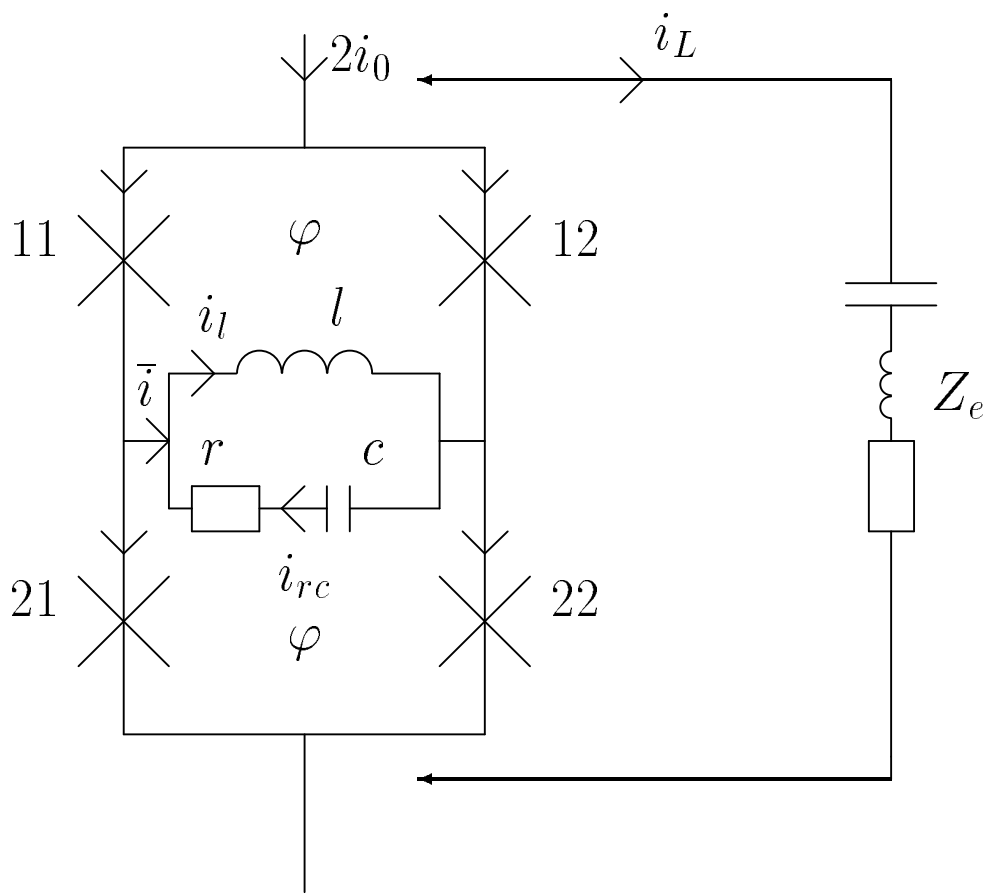
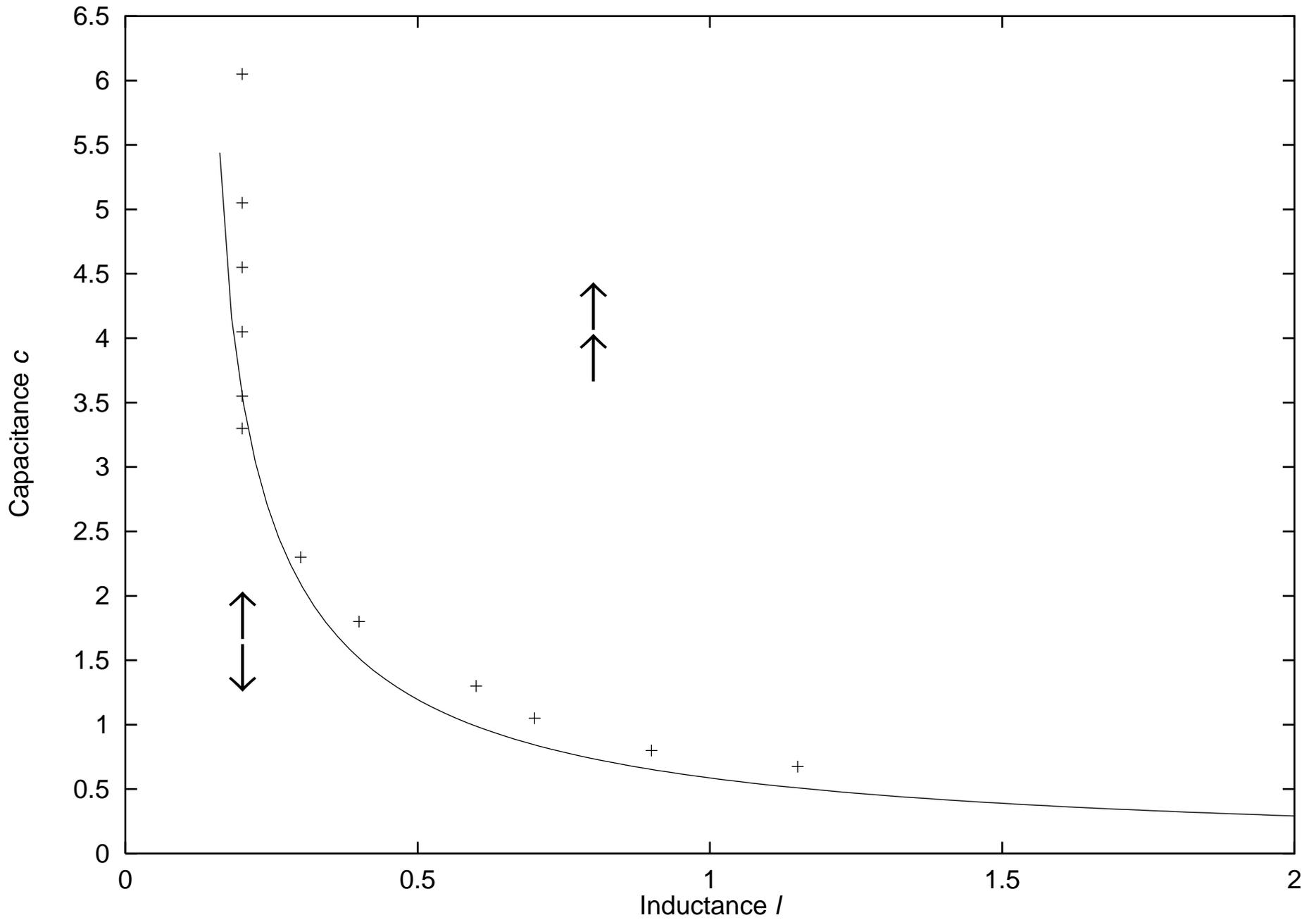
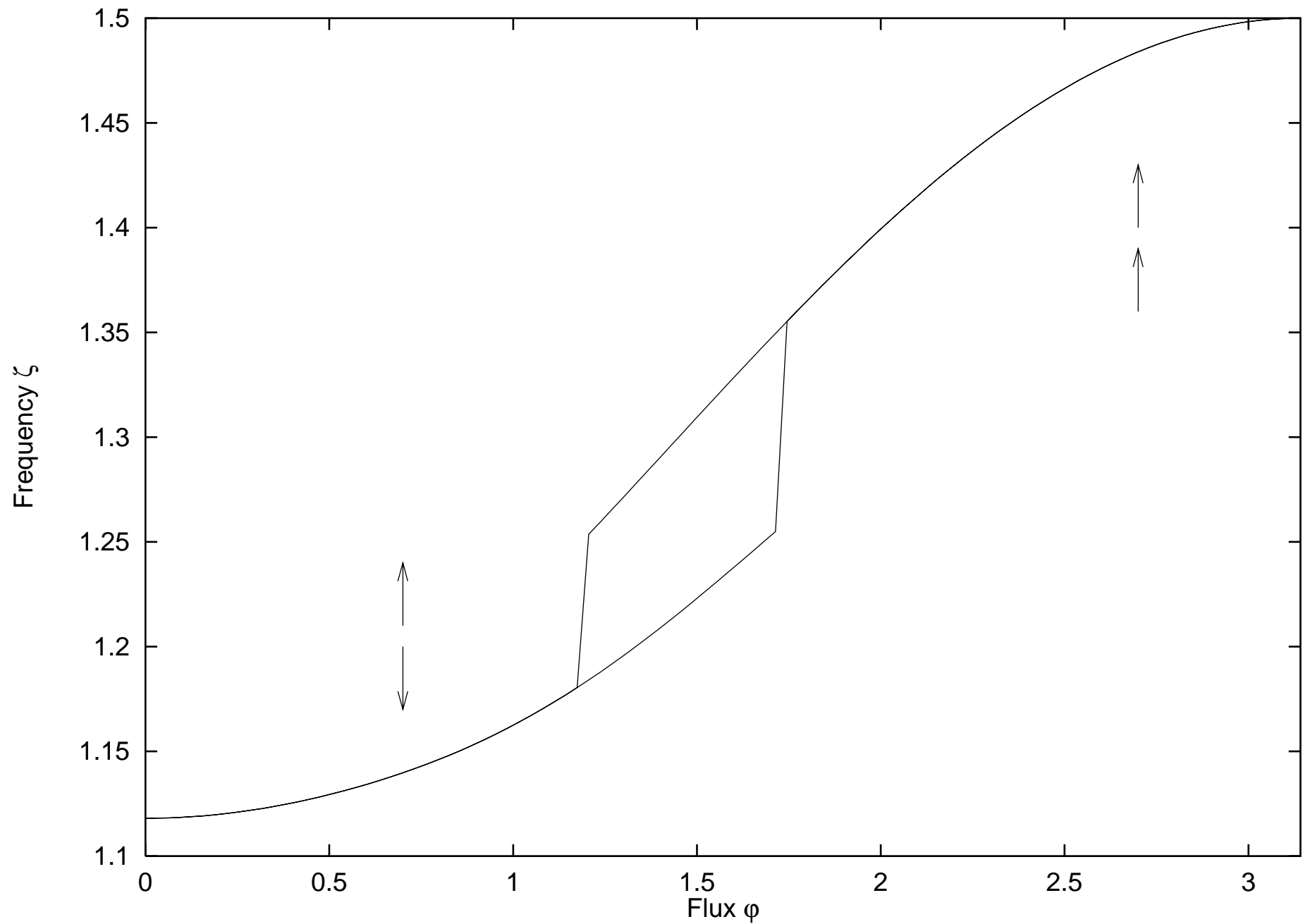
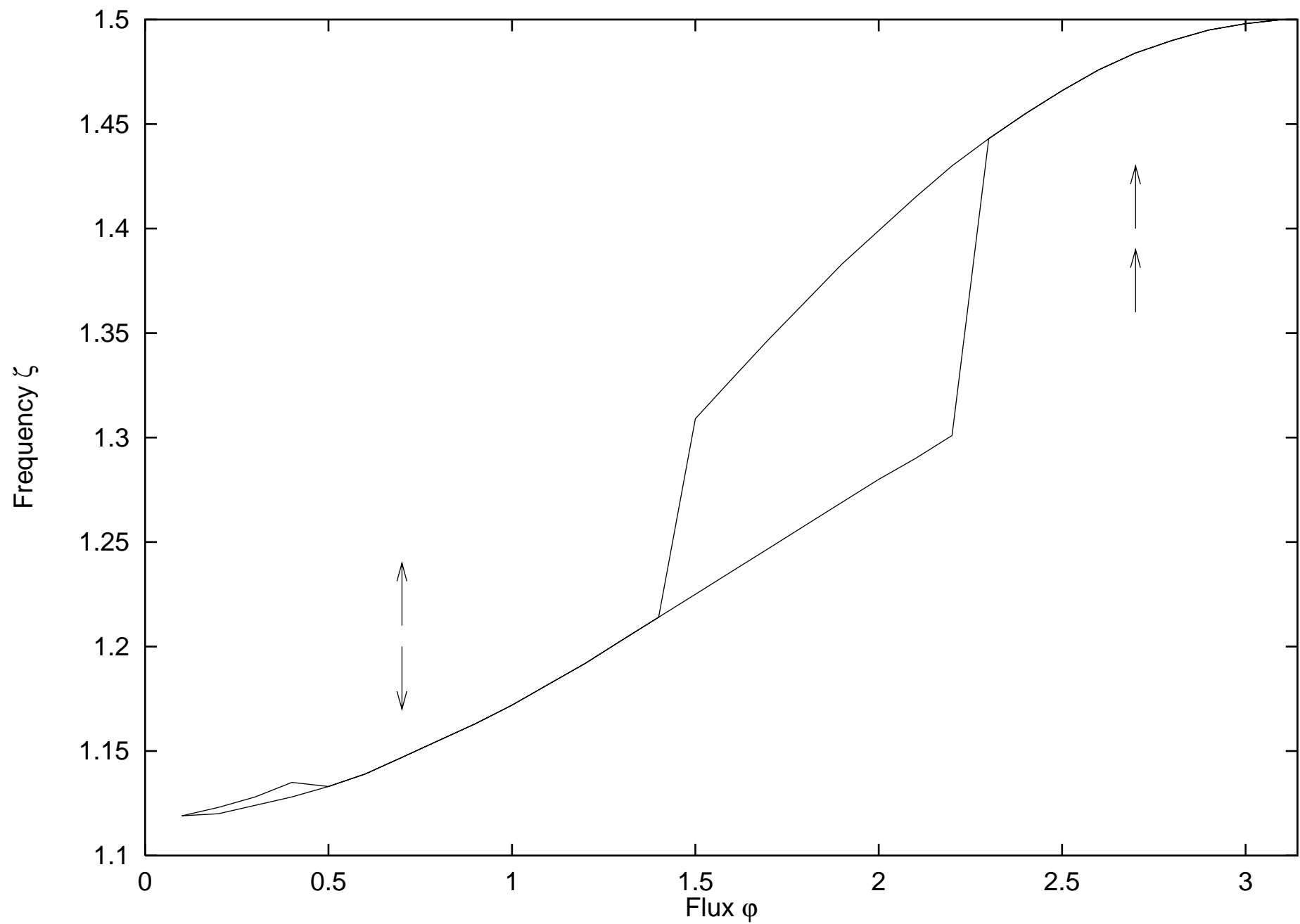
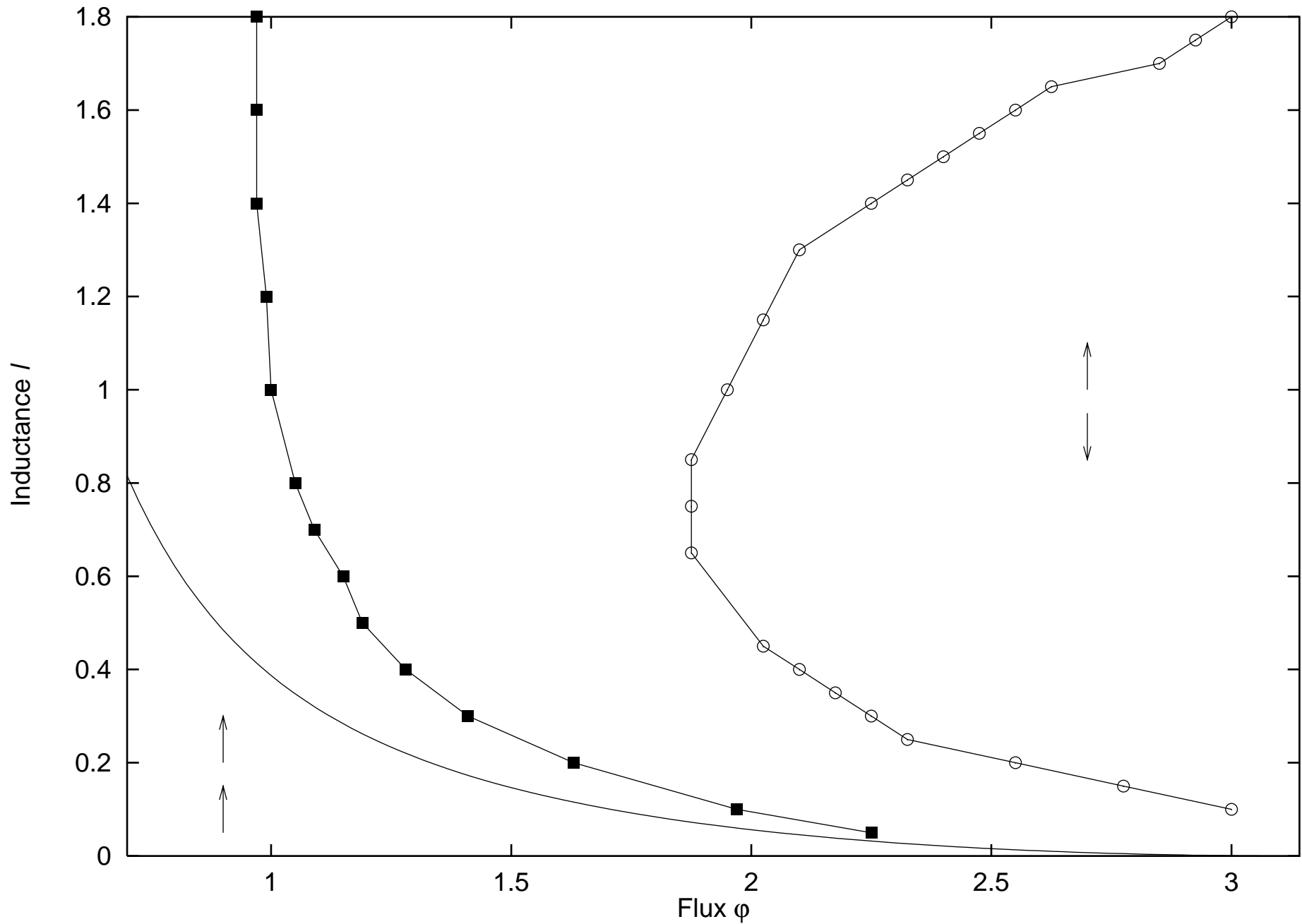
\end{document}